\documentclass[english]{article}
\usepackage{geometry}
\usepackage{graphicx}
\usepackage{subfigure}
\usepackage{amssymb}
\usepackage{bm}
\usepackage{fancyhdr}

\date{}

\usepackage{babel}

\title{Antiproton Flux in Cosmic Ray Propagation Models with Anisotropic Diffusion}

\author{Phillip Grajek$^{1}$, Kaoru Hagiwara$^{1,2}$\\
\textit{\small ${}^{1}$ KEK Theory Center, Tsukuba, 305-0801 Japan}\\
\textit{\small ${}^{2}$ Sokendai, Tsukuba, 305-0801 Japan}\\
}

\begin{document}

\maketitle

\abstract{
Recently a cosmic ray propagation model has been introduced, where anisotropic diffusion is used as a mechanism to allow for $\mathcal{O}(100)$ km/s galactic winds.  This model
predicts a reduced antiproton background flux, suggesting an excess is being observed.  We implement this model in GALPROP v50.1 and perform a $\chi^2$ analysis for B/C, $^{10}$Be/$^{9}$Be, and the recent PAMELA $\bar{p}/p$ datasets.  By introducing a power-index parameter $\alpha$ that dictates the dependence of the diffusion coefficient $D_{xx}$ on height $|z|$ away from the galactic plane, we confirm that isotropic diffusion models with $\alpha=0$ cannot accommodate high velocity convective winds suggested by ROSAT, while models with $\alpha=1$ ($D_{xx}\propto |z|$) can give a very good fit.  A fit to B/C and $^{10}$Be/$^{9}$Be data predicts a lower   $\bar{p}/p$ flux ratio than the PAMELA measurement at energies between approximately 2 GeV to 20 GeV.  A combined fit including in addition the $\bar{p}/p$ data is marginal, suggesting only a partial contribution to the measured antiproton flux.}

\renewcommand{\baselinestretch}{2}	
\large    

\renewcommand{\headrulewidth}{0.0pt}
\thispagestyle{fancy}
\rhead{KEK-TH-1423}

\section{Introduction}

In 2008-2009, the cosmic ray (CR) observation experiments PAMELA
\cite{Adriani:2008zr} and FERMI \cite{Abdo:2009zk} both recorded an anomalous
positron flux in excess of the expected astrophysical background.  Similar
excesses were also reported by the ATIC \cite{:2008zzr} and HESS
\cite{Aharonian:2009ah} experiments at different energies.  These observations
initiated a flurry of activity aimed at explaining the discrepancy.  Proposed
solutions include both astrophysical sources, such as local pulsars
\cite{Hooper:2008kg, Fujita:2009wk, Profumo:2008ms}, as well as more exotic
sources such as annihilating or decaying dark matter particles
 \cite{Feng:2010gw, Ishiwata:2010am, Ibarra:2008jk}.  The amount to which the latter contributes is extremely
important information, as many models of new physics that extend the Standard
Model predict the existence of particles that can make excellent candidates for
dark matter.  

Inherent to any of the possible solutions is an assumption for the underlying
physics describing how the CR particles are transported from their sources to
Earth.  At present, our understanding of how CR propagate throughout the galaxy
is mature but still incomplete.  The most realistic description of CR
propagation is currently obtained from models that assume CR transport occurs
via a combination of spatial diffusion (resulting from interaction with the
random and complicated galactic magnetic field) and convection (resulting from
interaction with large-scale streaming of CR away from the galactic disk). 
Explicit implementations have been presented in the literature.  These include
both semi-analytic \cite{Putze:2010zn, Maurin:2001sj} as well as fully numerical
treatments \cite{Strong:1998pw,Strong:2009xj, Evoli:2008dv, DiBernardo:2009ku}.

In the canonical propagation scenario, the diffusion behavior is assumed to be
identical everywhere in the galaxy: both in the gaseous disk as well as the
surrounding region known as the halo.  This isotropic diffusion model is capable
of reproducing numerous astrophysical observations well, and this has led to its
adoption as the model from which astrophysical backgrounds are normally derived.
 However, in isotropic diffusion models the convective wind velocity is limited
to no more than about 10-20 km/s.  Assuming a higher wind velocity results in
the incorrect prediction of several existing observations, most notably
measurements of secondary-to-primary ratios such as the boron-to-carbon (B/C)
ratio, as well as the ratio of radioactive beryllium isotopes ($^{10}$Be/$^9$Be)
\cite{Strong:2007nh, Maurin:2002hw, jones:1}. These ratios, in particular, are
highly sensitive to the parameters that control the various CR transport
mechanisms, and are often used to benchmark predictions of a given model.  
While very high velocity galactic winds on the order of 1000 km/s have been
observed in other galaxies \cite{Veilleux:2005ia}, until recently it was thought
that the Milky Way did not exhibit a wind, or that the wind velocity was limited
to very low values.  However, in 2007 the ROSAT satellite observed x-ray
emission from within the Milky Way that is consistent with the presence of a
galactic wind having a velocity of a few hundred km/s
\cite{Breitschwerdt:2008na, Everett:2007dw}. The measurements are also
consistent with a wind velocity profile that exhibits spatial dependence, and
follows the radial distribution of supernova remnants (SNR) in the galaxy.  The
presence of high-velocity, radially dependent galactic winds appear to be
natural and may also explain the large bulge-to-disk ratio observed by INTEGRAL
\cite{2005A&A...441..513K, 2007ESASP.622...25W}.  

Recently, a propagation model has been introduced that employs both a radially
dependent convective wind profile and \textit{anisotropic} diffusion, where the
diffusion behavior is not uniform throughout the galaxy, but instead varies as a
function of position \cite{Gebauer:2009hk}.  Specifically, the model assumes
that the intensity of CR diffusion increases with distance from the galactic
plane. This model is capable of supporting a convective wind velocities of
several hundred km/s, which is consistent with the ROSAT measurement, while
simultaneously reproducing benchmark measurements such as the B/C and
$^{10}$Be/$^9$Be ratios with similar accuracy to the traditional isotropic
diffusion model.  A crucial prediction of this scenario is that the
astrophysical anti-proton background flux appears lower than what is predicted
by isotropic diffusion models \cite{deBoer:2009tz}. Such an outcome has
significant implications for the indirect detection of particle dark matter, as
it would reduce constraints on models for new physics that tend to over-predict
the antiproton flux, assuming no excess is currently observed in the data
\cite{Donato:2008jk}.  

In this paper we implement the anisotropic propagation model of
\cite{Gebauer:2009hk} in the public CR propagation code GALPROP v50.1
\cite{Strong:2009xj}, and explore varying levels of diffusion anisotropy by
adjusting the rate at which the diffusion intensity increases with distance from
the galactic plane.  We perform a $\chi^2$ analysis for B/C, $^{10}$Be/$^{9}$Be
and the recent PAMELA $\bar{p}/p$ data and find the models that fit the data
best all exhibit spatially-dependent, anisotropic diffusion.  Further, we find
these models predict a $\bar{p}/p$ flux that is significantly lower than the
observed flux between 2 GeV and 20 GeV.  When we attempt a combined fit
including the $\bar{p}/p$ data, we find an increase in $\chi^2_{\mathrm{min}}$, by
$\Delta \chi^2=35.4$ for the 23 PAMELA data points considered.

\section{Anisotropic Diffusion} \label{sec:aniso}

As CR propagate they scatter and diffuse via interaction with turbulent
fluctuations in the galactic magnetic field.  Models of CR propagation
traditionally assume diffusion is isotropic throughout the propagation volume. 
This implies a constant density of scattering centers throughout the volume, and
an abrupt transition to free space (no scattering) at the outer boundary.  In
this case a propagating CR experiences, for a given momentum, the same
scattering mean free path in the gaseous disk as it does far into the halo.  In
the presence of a galactic wind flow driving CR away from the gaseous disk and
into the halo, the probability for a CR particle to return to the disk drops
rapidly with increasing distance into the halo, as it not only must survive the
convective action of the wind, but also an increasing number of scattering
events on its return trip.  The interplay between diffusive scattering and
convective transport strongly affects the times a CR particle spends in the
gaseous disk and halo regions of the galaxy.  This, in turn, affects the
prediction for the CR flux at Earth. In isotropic diffusion models the
convective wind speed is limited to less than approximately 20 km/s before it is
no longer possible to reproduce key observations such as the B/C ratio
\cite{Strong:2007nh, Maurin:2002hw, jones:1}.  This velocity limit is in
conflict with measurements from the ROSAT satellite, which are consistent with
galactic wind speeds up to several hundred km/s \cite{Breitschwerdt:2008na,
Everett:2007dw}.

In the anisotropic propagation model of \cite{Gebauer:2009hk}, CR diffusion is
taken to be spatially dependent, and increases monotonically with distance above
the gaseous disk.  This corresponds to a gradual reduction in the density of
scattering centers (and a smooth transition to free space) as CR travel from
the disk towards the outer boundary of the propagation volume.  In this scenario,
the mean scattering length \textit{increases} with increasing distance into the
halo.  Therefore, a CR particle in the halo that scatters back towards the disk
can travel further before experiencing its next scattering event, which is
likely in a region with a higher density of scattering centers.  This "trapping"
mechanism induces a drift of CR particles from areas of high diffusion to areas
with low diffusion, which assists in counteracting the outward convection of CR
due to the galactic wind.  In this way, convection velocities of several hundred
km/s can be accommodated while simultaneously preserving the CR residence times
in both the disk and halo required to reproduce the CR fluxes observed at Earth.

\section{Cosmic Ray Transport Model} \label{sec:analysis}

For our study, we utilize the CR propagation code GALPROP v50.1
\cite{Strong:2009xj}. GALPROP assumes that cosmic ray propagation in our galaxy
is governed by the following transport equation \begin{eqnarray} \label{diffeq}
\frac{\partial\Psi}{\partial t} - q(p,\bm{r},t) & = &
\bm{\nabla}\cdot(D_{xx}\bm{\nabla}-\bm{V})\Psi+\frac{\partial}{\partial
p}p^{2}D_{pp}\frac{\partial}{\partial
p}\frac{1}{p^{2}}\Psi\label{eq:transport}\\ &  & -\frac{\partial}{\partial
p}\left[\dot{p}\Psi-\frac{p}{3}(\bm{\nabla}\cdot\bm{V})\Psi\right]\nonumber \\ &
& -\frac{1}{\tau_{f}}\Psi-\frac{1}{\tau_{r}}\Psi\nonumber \end{eqnarray}  where 
$\Psi(p,\bm{r},t)$ is the density of CR with total momentum $p$ at location
$\bm{r}$ at time $t$. The cosmic ray source distribution, as well as
contributions from nuclear spallation and decays are included in
$q(p,\bm{r},t)$.  On the RHS, $D_{xx}$ is the spatial diffusion coefficient
while $\bm{V}(\bm{r})$ is the velocity profile of an assumed galactic wind.  
The term proportional to $D_{pp}$ accounts for random, stochastic accelerations
(reacceleration) that result from scattering, and is equivalent to diffusion in
momentum space. The term $p(\bm{\nabla}\cdot \bm{V})/3$ accounts for adiabatic
loss or gain in momentum, in the case that the wind velocity is location
dependent.  Finally, the term with $\dot{p}\equiv dp/dt$ accounts for momentum
losses such as inverse Compton scattering, bremsstrahlung, as well as
synchrotron radiation, while terms with $\tau_f^{-1}$ and $\tau_{r}^{-1}$
account for losses due to fragmentation and radioactive decay, respectively.
GALPROP assumes a cylindrical propagation region having maximum radius from the
galactic center $r=r_{\mathrm{max}}$, and maximum extension above and below the
galactic disk $z=L$.   Free escape ($\Psi=0$) is assumed at the boundaries of
the cylinder.

\subsection{GALPROP Modifications}

In this work the original GALPROP source code is modified in order to simulate
anisotropic diffusion and a radially-dependent, ROSAT compatible convective
wind. We follow closely the setup outlined in \cite{Gebauer:2009hk}. GALPROP
utilizes the Crank-Nicholson implicit method \cite{CNPress} to numerically solve
Eq.~(\ref{diffeq}) and it is necessary to compute the finite-difference
expansion coefficients in order to enact any modifications.  The coefficients
resulting from the inclusion of spatially dependent diffusion and convection
have been computed and they agree to those in \cite{Gebauer:2009hk} in the limit
of an equidistant computational grid.  We assume the following grid spacing:
$\Delta r=1$ kpc and $\Delta z=0.1$ kpc throughout this analysis. The
propagation cylinder height $L$ is allowed to vary, however the maximum radius
is fixed at $r_{\mathrm{max}}=20$ kpc.  

The spatial diffusion coefficient $D_{xx}$ is taken to be isotropic within a
slab of thickness 1 kpc above and below the galactic plane, and anisotropic at
distances $|z|\ge1$ kpc: \begin{eqnarray} D_{xx}(\rho)=\beta D_0
\left(\frac{\rho}{\rho_0}\right)^\delta, && |z|<1\;\mathrm{kpc}, \\ D_{xx}(\rho, z)=\beta
D_0 \left(\frac{\rho}{\rho_0}\right)^\delta |z|^\alpha, && |z|\ge1\;\mathrm{kpc},
\end{eqnarray}  where $\beta \equiv v/c$, $D_0$ is a constant, $\rho$ is the
particle rigidity (momentum/electric charge), and $\delta$ is the associated
power index. The reference rigidity $\rho_0$ is set to 4
GV. We
introduce a power index $\alpha$ which governs the diffusion gradient;
$\alpha=1$ was assumed in \cite{Gebauer:2009hk} and $\alpha=0$ reproduces the
original GALPROP parameterization.  The diffusion coefficient is related to the reacceleration coefficient
$D_{pp}$ via \begin{eqnarray} D_{pp}D_{xx}=\frac{4}{3}\frac{p^2
v_A^2}{\delta(4-\delta^2)(4-\delta)} \label{eq:reacc} \end{eqnarray}where $v_A$
is the speed of propagating Alfv\'en waves: weak disturbances in the magnetic
field that contribute to turbulent scattering. For the CR source distribution we
adopt the parameterization of Case and Bhattacharya \cite{Case:1998qg}:
\begin{equation} Q(r,z)=\left(\frac{r}{r_0}\right)^{1.69} exp \left[-3.33
\frac{r-r_0}{r_0} -\frac{|z|}{z_s}\right] \end{equation} where $r_0=8.5$ kpc is
the galacto-centric distance to Earth, and $z_s=200\;pc$.  The distribution has
a peak at $r\approx 0.51r_0$. The galactic wind velocity is taken to be
proportional to the source distribution at $z=0$ \begin{eqnarray} |V(r,z)|=Q(r,z=0)(V_0 +
|z|\cdot dV/dz). \label{eq:wind}   \end{eqnarray} We fix the constant velocity
component at $V_0=100$ km/s, and the gradient at $dV/dz=35$ km/s/kpc, which
models the galactic wind velocity profile indicated by the ROSAT data. In
GALPROP the galactic wind flow is one-dimensional along $\pm\bm{\hat{z}}$ and
always outward from the galactic plane. The cosmic ray injection spectrum
follows a power law in momentum,  $p^{-\eta}$.  For nuclei, $\eta=1.8$ for
rigidity $\rho<9$ GV, and $\eta=2.4$ for $\rho>9$ GV. Similarly, for electrons
$\eta=1.6$ for $\rho<4$ GV, 2.54 for $4\;\mathrm{GV}\le\rho\le100\;\mathrm{GV}$,
and 5.0 for  $\rho>100$ GV. We assume a constant $X_{\mathrm{CO}}$ factor
(conversion factor from CO integrated temperature to H$_2$ column density) of
$1.9\times10^{20}$, and an H/He ratio of 0.11.

\section{Analysis}

Using these settings as a kernel, we scan parameters $\alpha$, $D_0$, $v_A$, $\delta$,
and $L$ on the following grid:  $0\le \alpha \le 1$ in steps of 0.2, $4\;\mathrm{cm}^2/\mathrm{s}\le D_0
\le 6\;\mathrm{cm}^2/\mathrm{s}$ in steps of 0.2 $\mathrm{cm}^2/\mathrm{s}$,
$30\;\mathrm{km/s}\le v_{A}\le 60\;\mathrm{km/s}$ in steps of 2 km/s, $0.2\le
\delta \le 0.5$ in steps of 0.02, and $4\;\mathrm{kpc} \le L \le 8\;\mathrm{kpc}$ in
steps of 1 kpc. For each model point
$\{\alpha, D_0, v_A, \delta, L\}$ of the scan we compute the overall $\chi^2$
for the resulting B/C,  $^{10}$Be/$^{9}$Be, and $\bar{p}/p$ flux ratio curves.
For our study we assume \begin{eqnarray} \chi^2 =
\sum_{j=1}^M\sum_{i=1}^{N_j}\frac{(\bar{f}_i-f_{ij})^2}{\sigma_{ij}^2}
\label{eq:chisq} \end{eqnarray} where index $j$ runs over all $M$ experimental
datasets, each containing $N_j$ measurements $f_{ij}\pm\sigma_{ij}$ ($i=1$ to
$N_j$), and $\bar{f}_i$ is the computed flux evaluated at the specific energy of
the measurement $i$. Flux ratios are extracted using the CPLOT routine available
on the GALPROP
website\footnote{http://galprop.stanford.edu/web\_galprop/galprop\_home.html}. 
For comparison with data, the local interstellar (LIS) flux was corrected for
solar modulation using the ``force field'' approximation
\cite{1968ApJ...154.1011G}.  We assume the following modulation potentials, for
B/C, $^{10}$Be/$^{9}$Be, and $\bar{p}/p$ flux ratios, respectively, $\Phi=300$
MV, 350 MV, and 600 MV.    We include the following data.  For B/C: HEAO-3
\cite{Engelmann:1990zz}, CREAM \cite{2008APh....30..133A}, ACE
\cite{2000AIPC..528..421D}, and ATIC-2 \cite{Panov:2007fe}.  For
$^{10}$Be/$^9$Be: ISOMAX \cite{Hams:2004rz}, and ACE \cite{2001ApJ...563..768Y}.
For $\bar{p}/p$: PAMELA 2010 \cite{:2010rc}.  Except for the PAMELA $\bar{p}/p$
measurements, all the data are obtained from the Galactic Cosmic Ray Database
\cite{Strong:2009xp}.  In 4 of the 23 PAMELA data points the reported uncertainty
is asymmetric.  For those points  the larger error bar is used.  No correlation
among errors have been taken into account, as indicated by expression~(\ref{eq:chisq}). 
Our exercise should therefore be considered as a first
exploratory study of the allowed range of the parameters of the anisotropic
diffusion model of CR propagation in our galaxy.

\subsection{Combined fit to B/C and $^{10}$Be/$^9$Be}

We first attempt a combined fit to both the B/C and $^{10}$Be/$^9$Be datasets. We
find $\chi^2_{\mathrm{min}}/n_{\mathrm{dof}}=38.3/31$ within the
range of the parameter space explored. The best-fit values and corresponding
$3\sigma$ bounds for parameters $\alpha$, $D_0$, $\delta$, and $v_A$ are given in Table
~\ref{tab:ranges}. A bound for $L$ is not provided because the fit does not show
significant dependence on the propagation cylinder height.  

Within the $0\le \alpha \le 1$ range of our scan, the $\chi^2_{\mathrm{min}}$
occurs at the $\alpha=1$ boundary.  The corresponding $3\sigma$ allowed range
is $\alpha > 0.27$. The
isotropic diffusion limit $\alpha=0$ is excluded at $4\sigma$, which confirms
the finding of \cite{Gebauer:2009hk}. 
We remark that while our study focuses on the $0 \le \alpha \le 1$ region
of parameter space, a preliminary scan using a very coarse grid indicates
that $\chi^2$ increases with $\alpha$ for $\alpha > 1$.  A detailed examination
of the $\alpha > 1$ region is left for a future analysis.

The $3\sigma$
allowed range of the diffusion constant $D_0$ is
$4.3\times10^{28}\;\mathrm{cm}^2/\mathrm{s}\le D_0 \le
5.1\times10^{28}\;\mathrm{cm}^2/\mathrm{s}$, while that of the diffusion index
$\delta$ is $0.35\le \delta \le 0.43$.  These values are consistent with
previous studies involving GALPROP \cite{Lionetto:2005jd, Strong:1998pw} (see
also \cite{Strong:2007nh}) and also with magneto-hydrodynamic theory, which
predicts $\delta=1/3$ for Kolmogorov-type turbulence power spectrum, and
$\delta=1/2$  for a Kraichnan-type spectrum. Similarly, the $3\sigma$ range for
the Alfv\'en velocity $v_A$ is $47\;\mathrm{km/s} \le v_A \le 58\;
\mathrm{km/s}$.  These values are somewhat high compared to $v_A\simeq 30$ km/s
obtained for models with no convective wind \cite{Strong:1998pw}, and
$v_A\simeq20$ km/s expected for the warm ionized phase of the interstellar
medium \cite{1994ApJ...431..705S}.  This increase is due to the large velocity
gradient, $dV/dz=35$ km/s/kpc, assumed for the galactic wind (\ref{eq:wind}).  A
positive velocity gradient implies an adiabatic energy loss in response to the
reduction in pressure of the CR fluid (term proportional to
$p(\bm{\nabla}\cdot\bm{V})/3$ in  Eq.~(\ref{diffeq})).  The result is a
redistribution of the B/C flux spectra towards lower energies.   In order to
reproduce the distinct peak in B/C at $\sim 1$ GeV, this reduction is
compensated for with increased reacceleration, which is a mechanism that shifts
CR from lower energy to higher energy.  From Eq.~(\ref{eq:reacc}) this
translates into larger $D_{pp}$, or equivalently magnetic field disturbances
that propagate with larger Alfv\'en velocity ($v_A$).

\begin{figure}
\centering
\includegraphics[scale=1.1]{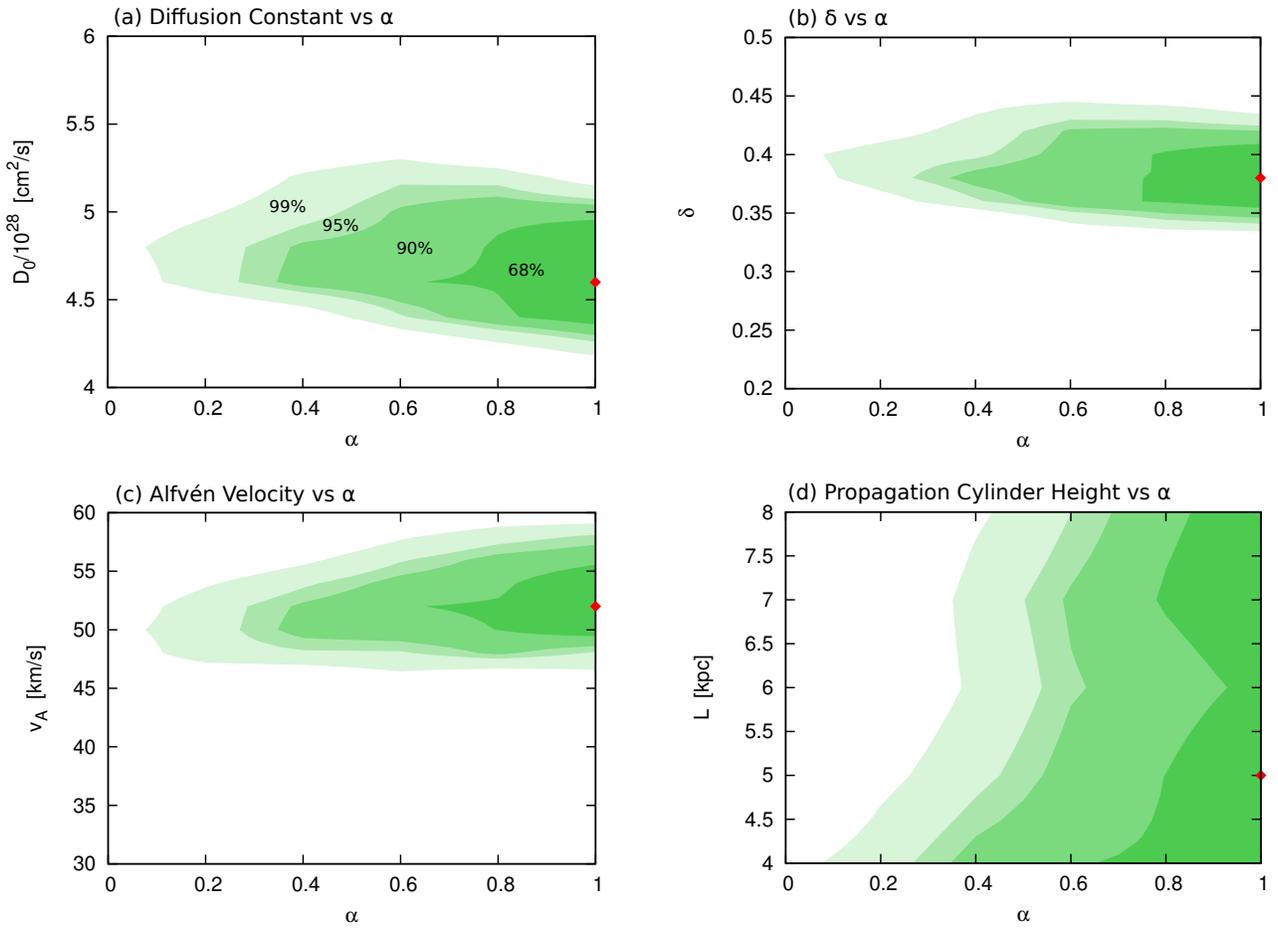}
\caption{Two-parameter likelihood surfaces for the transport parameters as a function of the diffusion gradient index $\alpha$, obtained for combined fit to B/C and $^{10}$Be/$^{9}$Be flux ratios.  Contours for $\Delta \chi^2=$ 2.3, 4.61, 5.99, and 9.21 are given, corresponding to
the 68\%, 90\%, 95\% and 99\% confidence level regions, respectively. Subfigures: (a) diffusion constant $D_0$, (b) diffusion coefficient power index $\delta$, (c) Alfv\'en velocity $v_A$, (d) propagation cylinder maximum height $L$.  The red diamond indicates the location of $\chi^2_{\mathrm{min}}$.  Whereas isotropic diffusion models with $\alpha=0$ ($D_{xx}$ independent of position) cannot accommodate the high velocity convective winds suggested by ROSAT, models with $\alpha=1$ ($D_{xx}\propto |z|$) can give a very good fit. }
\label{fig:CLrangesB}
\end{figure}

Figure~\ref{fig:CLrangesB} shows the two-parameter likelihood surfaces for
transport parameters $D_0$, $\delta$, $v_A$, and $L$ as a function of $\alpha$.
Contours for $\Delta \chi^2=$ 2.3, 4.61, 5.99, and 9.21 are given, corresponding
to the 68\%, 90\%, 95\% and 99\% confidence level regions, respectively. The
best-fit model point is indicated with a red diamond. Compared to models with
isotropic diffusion ($\alpha = 0$), those with spatially dependent, anisotropic
diffusion ($\alpha>0$) provide significantly better fits to the data. The
constraint on the propagation cylinder height $L$ (Fig.~\ref{fig:CLrangesB}d)
degrades with increasing $\alpha$.  As the diffusion gradient steepens, the
distance at which a CR transitions into what is effectively free space (no
scattering) occurs increasingly closer to the galactic plane, and away from the
assumed boundary of the propagation region at $L$.   This reduces the
sensitivity of the model to the precise value of $L$, which no longer defines
the extent of the magnetic corona, but becomes instead a computational
parameter.  In isotropic models, $L$ strongly controls the residence time of CR
in the halo, as its position determines the location of the free escape
boundary.  

\begin{figure}
\centering
\subfigure[]{\includegraphics[scale=1.3]{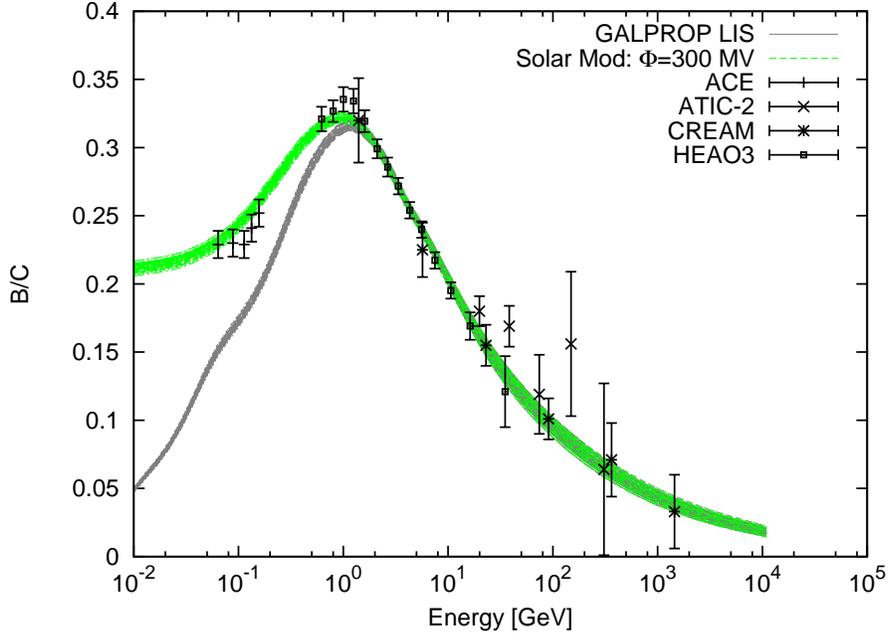} \label{fig:bcfitB}}
\subfigure[]{\includegraphics[scale=1.3]{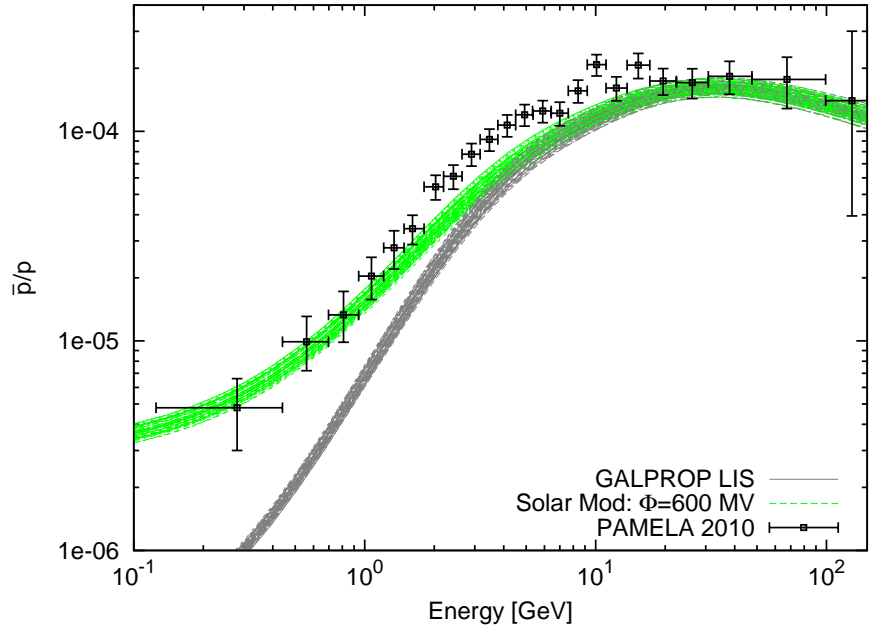} \label{fig:bcfitE} }
\caption{Predicted flux distribution for models within the 68\% confidence level region (four parameter fit $\Delta \chi^2=4.72$), for the combined fit to B/C and $^{10}$Be/$^{9}$Be. The grey (dark) curves indicate the local interstellar (LIS) flux, while the green (light) curves have been corrected for solar modulation using the ``force field`` approximation (modulation potential indicated). The B/C flux ratio is reproduced well, however the predicted $\bar{p}/p$ flux  is significantly lower than the PAMELA measurements between approximately 2 GeV to 20 GeV.}
\label{fig:fitsB}
 \end{figure}

Figure~\ref{fig:fitsB} shows the B/C  and $\bar{p}/p$ flux ratio curves for
model points that lie within the 68 \% CL region, assuming a four parameter
fit ($\Delta \chi^2=4.72$).  The grey (dark) lines correspond to the local
interstellar (LIS) flux, while the green (light) lines correspond to the flux
after correcting for solar modulation. The B/C flux is reproduced well. However,
the $\bar{p}/p$ flux is significantly lower than the PAMELA measurement at
energies between approximately 2 GeV and 20 GeV.

\subsection{Combined fit to B/C, $^{10}$Be/$^9$Be, and $\bar{p}/p$}

\begin{figure}
\centering
\subfigure[]{\includegraphics[scale=1.3]{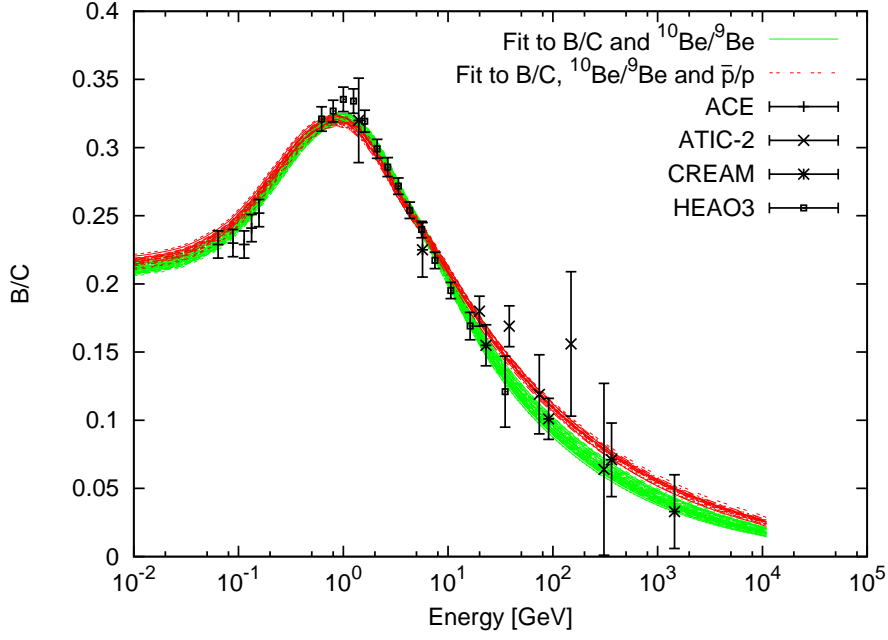} \label{fig:pbarfitB}}
\subfigure[]{\includegraphics[scale=1.3]{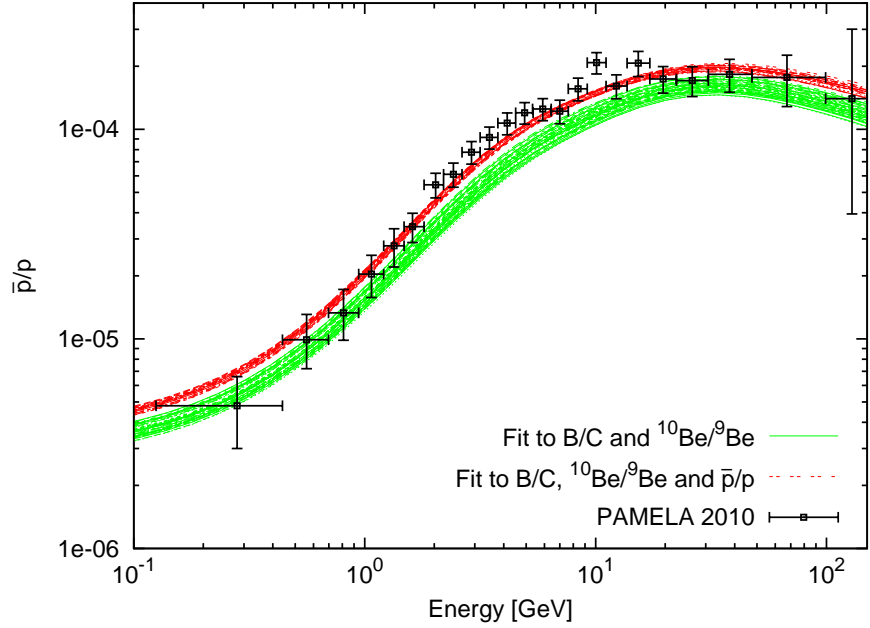} \label{fig:pbarfitE} }
\caption{The predicted flux distribution obtained for models within the 68\% confidence
level region for the combined fit to the B/C, $^{10}$Be/$^{9}$Be, and $\bar{p}/p$ data
in red (dark) superimposed on the results obtained for the fit to B/C and $^{10}$Be/$^{9}$Be 
only in green (light).  The $\chi^2_{\mathrm{min}}$ is observed to 
increase by $\Delta \chi^2=35.4$ for the additional 23 PAMELA data points considered. The fit to the B/C flux is degraded as the distribution
does not reproduce well the high precision HEAO-3 measurements above 10 GeV.  The
fit to $\bar{p}/p$ is marginally improved, however the distribution is still
lower than the PAMELA observation between approximately 2 GeV and 15 GeV.}
\label{fig:fitsE}
 \end{figure}

We next attempt a combined fit to the B/C, $^{10}$Be/$^9$Be, and also the
$\bar{p}/p$ data. We observe an increase in $\chi^2_{\mathrm{min}}$, by $\Delta
\chi^2=35.4$ for the 23 PAMELA data points considered. Figure~\ref{fig:fitsE}
shows the 68\% CL curves obtained for the combined fit in red (dark)
superimposed upon the results from Fig.~\ref{fig:fitsB} (B/C and
$^{10}$Be/$^9$Be only) in green (light).  The fit to the B/C data is now
degraded somewhat as the models do not reproduce well the high precision HEAO-3
data above 1 GeV.  The fit to the $\bar{p}/p$ data is marginally improved.
However, an excess flux remains between approximately 2 GeV and 15 GeV.

\begin{figure}
\centering
\includegraphics[scale=1.1]{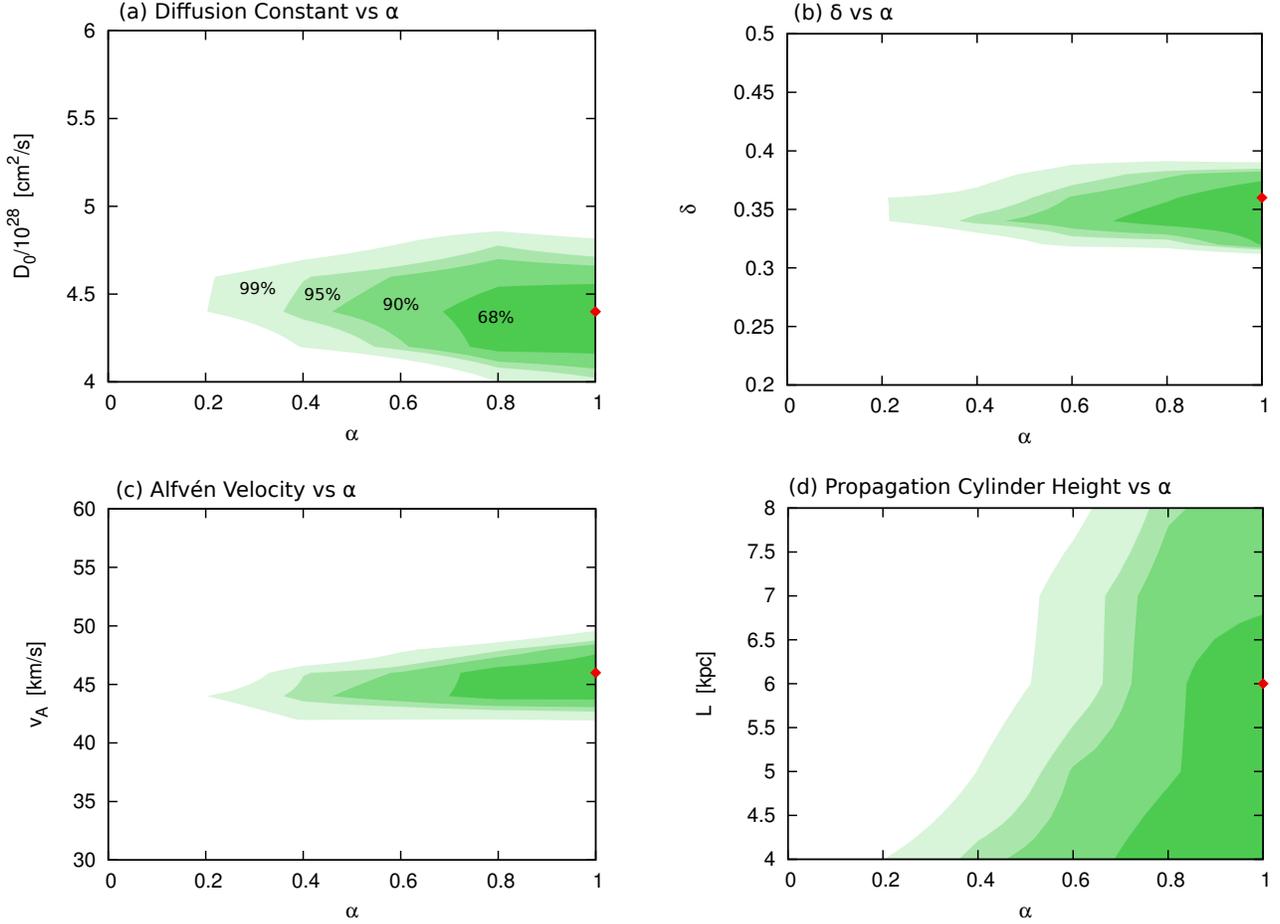}
\caption{Two-parameter likelihood surfaces for the transport parameters as a function of the diffusion gradient index $\alpha$, obtained for combined fit to B/C, $^{10}$Be/$^{9}$Be and $\bar{p}/p$ flux ratios.  The contours are as in Fig.~\ref{fig:CLrangesB}.  We find an increase of $\chi^2_{\mathrm{min}}$ by $\Delta \chi^2=35.4$ for the 23 PAMELA data points considered, and a significant reduction of the Alfv\'en wave velocity ($v_A$)}
\label{fig:CLrangesE}
 \end{figure}

The lower half of Table~\ref{tab:ranges} gives the best-fit parameter values and
corresponding $3\sigma$ allowed range.  The two-parameter likelihood surfaces
are shown in Fig.~\ref{fig:CLrangesE}.  Models with $\alpha > 0$ still provide
the best fits to the data, however overall larger values of $\alpha$ (steeper
diffusion gradient) are required: The best fit occurs at the $\alpha=1$ 
boundary of our scan range ($0\le \alpha \le 1$), and the $3\sigma$ allowed
range is $\alpha > 0.51$.  The allowed range for parameters $D_0$,
$v_A$, and $\delta$ are now shifted to lower values relative to those obtained
for the fit to B/C and $^{10}$Be/$^9$Be only, with the most pronounced reduction
occurring for the Alfv\'en velocity.  A reduction in Alfv\'en velocity corresponds
to weakening of the reacceleration mechanism (smaller $D_{pp}$ in
~(\ref{eq:reacc})), a process that boosts CR momentum to higher values.  Weaker
reacceleration causes both the B/C and $\bar{p}/p$ flux distributions to shift
to lower energies, which improves the fit to $\bar{p}/p$ but simultaneously
degrades the fit to B/C.  Lower values for $D_0$ and $\delta$ reduce the
intensity of spatial diffusion, which forces CR to remain longer in the gaseous
disk. This increases the rate of secondary production and boosts the B/C flux,
which helps to compensate for the reduction in Alfv\'en velocity. 

\section{Conclusions}

We consider the anisotropic propagation model of \cite{Gebauer:2009hk}, which
assumes that CR diffusion increases linearly with distance from the galactic
disk, and radially dependent convective wind flows with maximum velocity of
$\mathcal{O}(100)$ km/s suggested by the ROSAT measurement.  This model is capable of reproducing key observations such as the
B/C and Be/Be flux ratios, whereas traditional models based on isotropic
diffusion cannot support convective wind velocities of more than approximately
20 km/s.  

We implement this model in GALPROP, and introduce a new parameter, $\alpha$,
which controls the dependence of the diffusion coefficient $D_{xx}$ on the
distance $|z|$ above the gaseous disk; $\alpha=0$ corresponding to isotropic
diffusion, while $\alpha=1$ a linear increase as in \cite{Gebauer:2009hk}.  We
then perform a $\chi^2$ analysis for the B/C, $^{10}$Be/$^9$Be, and the recent
PAMELA $\bar{p}/p$ data sets in order to constrain the transport parameters.  For
a combined fit to the B/C and $^{10}$Be/$^9$Be flux only, we observe
$\chi^2_{\mathrm{min}}/n_{\mathrm{dof}}=38.3/31$, indicating a good fit to both
data sets.  We find $\alpha > 0.27$, $4.3\times10^{28}\;\mathrm{cm}^2/\mathrm{s}\le D_0 \le
5.1\times10^{28}\;\mathrm{cm}^2/\mathrm{s}$, $0.35\le \delta \le 0.43$, and
$47\;\mathrm{km/s} \le v_A \le 58\; \mathrm{km/s}$ at $3\sigma$ for the range of
parameter space we explored. The Alfv\'en velocity is higher than expected, and
due to a large wind velocity gradient, $dV/dz=35$ km/s/kpc, required to model
the wind profile suggested by the ROSAT measurements. The isotropic diffusion
case $\alpha=0$ is excluded at $4\sigma$.  Models within the 68\% CL region
predict a lower $\bar{p}/p$ flux than the PAMELA observation between 2 GeV to 20
GeV. For a combined fit to B/C, $^{10}$Be/$^9$Be and also the PAMELA $\bar{p}/p$
data we observe an increase in $\chi^2_{\mathrm{min}}$, by $\Delta \chi^2=35.4$ for the
additional 23 PAMELA data points considered, indicating the overall fit is
marginal. In this case we find $\alpha > 0.51$, $4.1\times10^{28}\;\mathrm{cm}^2/\mathrm{s}\le
D_0 \le 4.8\times10^{28}\;\mathrm{cm}^2/\mathrm{s}$, $0.33\le \delta \le 0.38$,
and $43\;\mathrm{km/s} \le v_A \le 49\; \mathrm{km/s}$ at $3\sigma$, which are
consistent with expectations.  Models within the 68\% CL region give an
improved fit to the $\bar{p}/p$ flux, however the predicted distributions are
still below the PAMELA observation between approximately 2 GeV to 15 GeV.

These results suggest that spatially dependent diffusion represents a viable
solution to the problem of accommodating $\mathcal{O}(100)$ km/s convective wind flows in
models of CR propagation, and that in this case the observed antiproton flux may
exhibit an excess above the astrophysical background.  An anomalous excess in
the antiproton flux would have many implications for the indirect detection of
particle dark matter (DM). For example,  pair annihilation of light neutralino
DM can explain the positron excess observed by e.g. PAMELA \cite{Grajek:2008pg}.
These annihilations necessarily produce hadrons in addition to
leptons, and so a contribution to the antiproton flux is also expected from this
scenario.

\pagebreak

\begin{table}[h]
\centering
\begin{tabular}{|c|c|c|c|}
\hline
Combined Fit & Parameter & Best Fit Value & 3$\sigma$ Range \\
\hline\hline
B/C, $^{10}$Be/$^{9}$Be & 
\begin{tabular}{c}
$\alpha$ \\ $D_0$ \\ $v_A$ \\ $\delta$ \\
\end{tabular} &
\begin{tabular}{c}
1.0 \\ $4.6 \times10^{28}$ cm$^2$/s \\ $52$ km/s \\ 0.38 \\
\end{tabular} &
\begin{tabular}{c}
$>$ 0.27 \\ 4.3 - 5.1 $\times10^{28}$ cm$^2/$s \\ 47 - 58 km/s \\ 0.35 - 0.43\\
\end{tabular}\\
\hline
B/C, $^{10}$Be/$^{9}$Be, $\bar{p}/p$ & 
\begin{tabular}{c}
$\alpha$ \\ $D_0$ \\ $v_A$ \\ $\delta$ \\
\end{tabular} &
\begin{tabular}{c}
1.0 \\ $4.4 \times10^{28}$ cm$^2$/s \\ 46 km/s \\ 0.36 \\
\end{tabular} &
\begin{tabular}{c}
$> 0.51$ \\ 4.1 - 4.8 $\times10^{28}$ cm$^2/$s \\ 43 - 49 km/s \\ 0.33 - 0.38 \\
\end{tabular} \\
\hline
\end{tabular}
\caption{Best-fit model parameters for the combined $\chi^2$ fits explored in this study. The 3$\sigma$ allowed range is given for $L=5$ kpc for the fit to B/C and $^{10}$Be/$^9$Be (above),
and for $L=6$ kpc for the fit including $\bar{p}/p$ (bottom), where we find the minimum $\chi^2$.
 No bound is provided for parameter $L$, as the fit does
not exhibit strong dependence on the propagation cylinder height.} \label{tab:ranges}
\end{table}

\end{document}